\documentclass[%
 reprint,
 superscriptaddress,
 amsmath,amssymb,
 aps,
 pra,
 longbibliography,
lengthcheck,%
]{revtex4-1}

\usepackage[english]{babel}
\usepackage{graphicx}
\usepackage{dcolumn}
\usepackage{bm}
\usepackage{color}
\usepackage{ulem}
\usepackage{hyperref}
\usepackage{braket}

\begin{document}

\preprint{APS/123-QED}

\title{
Field-induced metal-insulator transition, Chern insulators, and topological semimetals\\in a clean magnetic semiconductor GdGaI
}

\author{Kazuki Guzman}
\affiliation{
Department of Physics, Institute of Science Tokyo, Meguro, Tokyo, 152-8551, Japan
}

\author{Hiroaki Ishizuka}
\affiliation{
Department of Physics, Institute of Science Tokyo, Meguro, Tokyo, 152-8551, Japan
}

\date{\today}

\begin{abstract}
Non-coplanar magnetic order in low-carrier-density semiconductors provides a platform on which spin-charge coupling can reshape the electronic structure and induce nontrivial topological phases.
Motivated by the recent discovery of the four-sublattice triple-$q$ order in the magnetic semiconductor GdGaI, we study an effective theory that couples a Ga $4p$ hole pocket at the $\Gamma$ point to three Gd $5d$ electron pockets at the $M$ points through four exchange channels.
For the antiferromagnetic umbrella state with zero net magnetization, the model hosts trivial ($C = 0$) and $C = \pm 4$ Chern insulator phases separated by metallic regions; by deriving an analytical low-energy theory at the $\Gamma$ point, we show that the topological phase boundary is described by two degenerate double-Weyl semimetals, naturally explaining the $\Delta C = 4$ jump in the Chern number. 
In addition, a nodal-line-like state pinned near the Fermi level emerges in the absence of the $p$-$d$ exchange coupling, which separates the $C=\pm4$ phases for $\theta=\arccos(1/3)$ into two. 
Tuning the canting angle by an external magnetic field drives an insulator-to-metal transition out of the Chern insulator phase while leaving the trivial insulator largely intact, and stabilizes an additional $C = \pm 2$ Chern insulator phase when the uniform-magnetization exchange couplings become appreciable. 
These results identify GdGaI and its sister compounds as highly tunable platforms for realizing topological phases and field-induced metal-insulator transitions in clean magnetic semiconductors.
\end{abstract}

\pacs{
}

\maketitle

\section{Introduction}
%
%
Conducting magnets with non-coplanar spin configurations display distinctive electronic and transport properties rooted in nontrivial quantum-mechanical effects, and hold promise for technological applications.
A well-studied example is the anomalous Hall effect (AHE) driven by scalar spin chirality~\cite{Ye1999a,Ohgushi2000a,Taguchi2001a,Tatara2002a}.
This phenomenon originates from the spin Berry phase acquired by conduction electrons traversing a non-coplanar spin texture, and enables electrical detection of chiral magnetic states~\cite{Neubauer2009a,Takatsu2010a,Kanazawa2011a} as well as a sizable AHE in materials with small net magnetization~\cite{Nakatsuji2015a}.
In clean systems, non-coplanar magnetic orders can further produce a large AHE through skew scattering~\cite{Ishizuka2018a,Yang2020a,Fujishiro2021a,Ishizuka2021a}, and may even stabilize a topological quantum Hall state~\cite{Hamamoto2015a}.
To advance our understanding of these phenomena, it is desirable to identify candidate materials that combine a simple electronic structure with non-coplanar magnetic order.

\begin{figure}[tbp]
  \includegraphics[width=\linewidth]{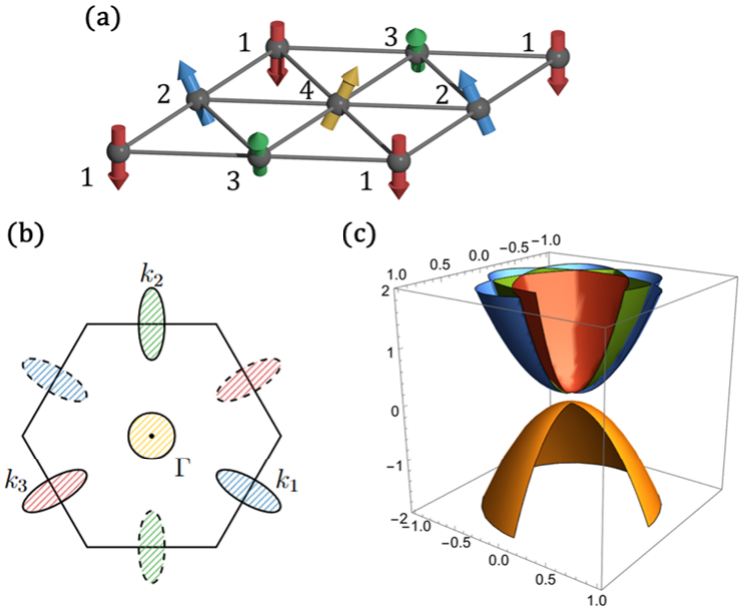}
  \caption{
  (a) A schematic of GaGdI with the four-sublattice order. (b) The electronic band of GaGdI. Three electron pockets are located at the $M$ points, and a hole pocket is at the $\Gamma$ point. (c) Electronic bands in the folded Brillouin zone.
  }\label{fig:model}
\end{figure}

A promising platform in this regard is the magnetic semiconductor GdGaI, which has recently been reported to exhibit a triple-$q$ four-sublattice order~\cite{Okuma2024a}.
In this magnetic structure, three spins point upward with a finite canting, and the fourth spin downward~\cite{Martin2008a,Akagi2010a,Kato2010a,Takagi2023a}, as illustrated in Fig.~\ref{fig:model}(a).
Because GdGaI is a low-carrier-density semiconductor hosting a non-coplanar magnetic state~\cite{Okuma2024a,Kaneko2026a,Vylet2026a}, it provides a particularly clean setting in which to study spin-charge coupling and its impact on the electronic properties.
Indeed, a recent experiment reported a large AHE in this material~\cite{Okuma2026a}, implying that the four-sublattice magnetic order substantially modifies the electronic structure.

%
%

In this work, we develop an effective $k \cdot p$ theory for GdGaI and investigate its electronic states and topological properties.
Building on first-principles calculations~\cite{Okuma2024a,Kaneko2026a}, we formulate a minimal model incorporating the four-sublattice order~\cite{Martin2008a}.
We find that this model hosts a rich phase diagram comprising $C=2$ and $C=4$ Chern insulators separated by a semimetallic phase.
By deriving a low-energy effective Hamiltonian from the $k \cdot p$ theory, we further identify this intervening state as a double-Weyl semimetal.
In addition, the application of a magnetic field is shown to drive a metal-insulator transition.
These results suggest that GdGaI and its sister compounds are highly tunable platforms with the potential to host a variety of topological phases and field-induced metal-insulator transitions.

The remainder of this paper is organized as follows. In Sec.~\ref{sec:model}, we introduce the $k \cdot p$ Hamiltonian and the four-sublattice magnetic order. Section~\ref{sec:results} presents the phase diagram, Hall conductivity, and Chern insulator phases. In Sec.~\ref{sec:JA-and-JB}, we derive the low-energy effective Hamiltonian and analyze the double-Weyl semimetallic state.
The field-induced metal-insulator transition is discussed in Secs.~\ref{sec:MItransition}-\ref{sec:metal}. Section~\ref{sec:summary} is devoted to the summary and discussions.

\section{Model}\label{sec:model} 

The first-principles calculations show that the electronic states near the Fermi level consist of the Ga $4p$ valence band at the $\Gamma$ point and Gd $5d$ conduction bands at the $M$ points (Fig.~\ref{fig:model}(b))~\cite{Okuma2024a,Kaneko2026a}.
To study how these bands deform in the presence of the triple-$q$ order, we study the electronic state of the following effective Hamiltonian $H=H_0+H_K$, where
\begin{align}
    H_0=&\sum_{\bm k,\sigma}\left[\alpha_0k^2-\Delta\right]p_{\bm k\sigma}^\dagger p_{\bm k\sigma}\nonumber\\
    +&\sum_{\substack{\bm k,\sigma,n}}\left[\alpha_x(\bm k\cdot \bm a_x^{(n)})^2+\alpha_y(\bm k\cdot \bm a_y^{(n)})^2+\Delta\right]d_{n\bm k\sigma}^\dagger d_{n\bm k\sigma}\nonumber\\
    -&\mu\sum_{\substack{\bm k,\sigma}}p_{\bm k\sigma}^\dagger p_{\bm k\sigma}-\mu\sum_{\bm k,\sigma,n}d_{n\bm k\sigma}^\dagger d_{n\bm k\sigma},
\end{align}
\begin{align}
    H_K=&J_A^0\sum_{\bm k,\alpha,\beta}\left[\bm S_{\bm 0}\cdot\bm\sigma\right]_{\alpha\beta}p_{\bm k\alpha}^\dagger p_{\bm k\beta}\nonumber\\
    +&J_B^0\sum_{\bm k,\sigma,n}\left[\bm S_{\bm 0}\cdot\bm\sigma\right]_{\alpha\beta}d_{n\bm k\alpha}^\dagger d_{n\bm k\beta}\nonumber\\
    +&J_A\sum_{\bm k,n,\alpha,\beta}\left[\bm S_{\bm k_n}\cdot\bm\sigma\right]_{\alpha\beta}\left(d_{n\bm k\alpha}^\dagger p_{\bm k\beta}+\text{h.c.}\right)\nonumber\\
    +&J_B\sum_{\substack{\bm k,\sigma,\alpha,\beta\\n\ne m}}\left[\bm S_{\bm k_n-\bm k_m}\cdot\bm\sigma\right]_{\alpha\beta}\left(d_{n\bm k\alpha}^\dagger d_{m\bm k\beta}+\text{h.c.}\right).\label{eq:HK}
\end{align}
Here, $p_{\bm k\sigma}^\dagger$ ($p_{\bm k\sigma}$) and $d_{n\bm k\sigma}^\dagger$ ($d_{n\bm k\sigma}$) are the creation (annihilation) operators for the $p$-orbital electrons at the $\Gamma$ point and the $d$-orbital electrons at the $n$-th $M$ point ($n=1,2,3$), respectively, with crystal momentum $\bm k$ and spin $\sigma = \uparrow,\downarrow$.
The parameter $\alpha_0$ is the inverse effective mass coefficient of the isotropic $p$ band, $\mu$ is the chemical potential, and $\Delta$ is the energy offset between the $p$ valence band and the $d$ conduction bands; $2\Delta$ corresponds to the band gap in the nonmagnetic state.
The $d$-band dispersion is anisotropic, reflecting the ellipsoidal shape of the Fermi pockets at the $M$ points; $\alpha_x$ and $\alpha_y$ are the inverse effective mass coefficients along the two principal axes of each pocket.
The principal axes for $n$th pocket is defined by $\bm a_x^{(n)}$ and $\bm a_y^{(n)}$, which is
\begin{align}
\bm a_{x}^{(1)}=(1/2,\sqrt3/2),\quad\bm a_{y}^{(1)}=(\sqrt3/2,-1/2),\label{eq:avec}
\end{align}
and the other two sets given by rotating Eq.~\eqref{eq:avec} by 120$^\circ$.

The spin-charge coupling $H_K$ describes the exchange interaction between the itinerant electrons and the localized magnetic moments.
Here, $\bm S_{\bm q}=(1/N)\sum_{\bm r}S(\bm r)e^{-i\bm k\cdot\bm r}$ denotes the Fourier component of the magnetic order with $N$ being the number of magnetic unit cells, and $\bm\sigma = (\sigma_x,\sigma_y,\sigma_z)$ with $\sigma_{x,y,z}$ being the Pauli matrices.
For the case of the four-sublattice orders with the magnetic supercell in Fig.~\ref{fig:model}, nonzero Fourier component is allowed only for $\bm k=\bm0$, $\bm k_1=\frac{2\pi}a(\frac1{2\sqrt3},-\frac16)$, $\bm k_2=\frac{2\pi}a(0,\frac13)$, and $\bm k_3=\frac{2\pi}a(\frac1{2\sqrt3},\frac16)$. 
In general, there are four different exchange couplings, whose coupling constants are $J_A^0$, $J_B^0$, $J_A$, and $J_B$, as given in Eq.~\eqref{eq:HK}. 

For the triple-$q$ umbrella state with canting angle $\theta$, $\bm S_{\bm q}$ reads
\begin{align}
&\bm S_{\bm 0}=(0,0,-1+3\cos\theta),\nonumber\\
&\bm S_{\bm k_1}=(-\sqrt3\sin\theta,\sin\theta,-1-\cos\theta),\nonumber\\
&\bm S_{\bm k_2}=(0,-2\sin\theta,-1-\cos\theta),\nonumber\\
&\bm S_{\bm k_3}=(\sqrt3\sin\theta,\sin\theta,-1-\cos\theta).
\end{align}
Note that, $\bm k_1=\bm k_3-\bm k_2$, $\bm k_2=\bm k_1-\bm k_3$, and $\bm k_3=\bm k_2-\bm k_1$.
Therefore, the $J_B$ term in Eq.~\eqref{eq:HK} appears in addition to the $J_A$ term.

In the following, we set the lattice parameter $a=1$ for the sake of simplicity, and the band parameters to $\alpha_0=-2.7$, $\alpha_x= 8.0$, $\alpha_y= 2.7$, and $\Delta=0.04$, unless noted otherwise.
The dispersion without $H_K$ is shown in Fig.~\ref{fig:model}(c); the two bottom bands are the Ga $4p$ bands and the other six bands are the Gd $5d$ bands. 
We fix the Fermi level so that the filling is $n=2$, the filling at which the bottom two bands are filled in the case of Fig.~\ref{fig:model}(c).

\begin{figure}[tbp]
  \includegraphics[width=\linewidth]{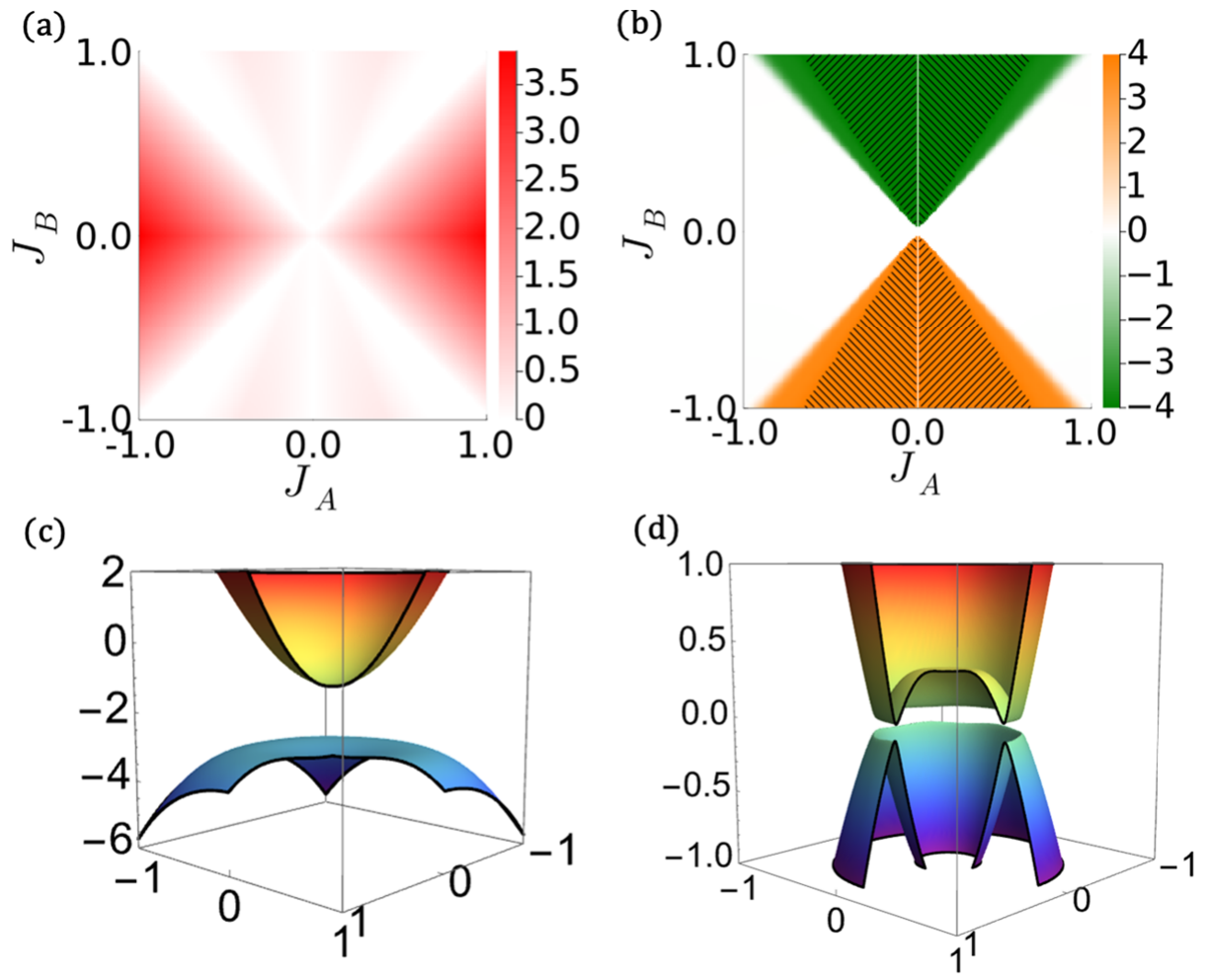}
  \caption{
  Contour plot of (a) the band gap and (b) Hall conductivity in the $J_A$-$J_B$ plane. The hatched regions are Chern insulators. Figures (c) and (d) are the band structures of the trivial ($J_A=-1.0$, $J_B=-0.5$) and $C=4$ Chern ($J_A=-0.1$, $J_B=-0.5$) insulators, respectively.
  The results are for $\alpha_0=-2.7$, $\alpha_x= 8.0$, $\alpha_y= 2.7$, $\Delta=0.04$, filling $n=2$, $\theta=\arccos(1/3)$, and $J_{A,B}^0=0$. 
  }\label{fig:AFM}
\end{figure}

\section{Results}\label{sec:results}

\subsection{Chern insulator and topological Hall effect in antiferromagnetic state} \label{sec:JA-and-JB}

Figure~\ref{fig:AFM} shows the phase diagram for the AFM umbrella state with zero net magnetization, i.e., $\theta=\arccos(1/3)\simeq0.39\pi$ ($\cos(\theta)=1/3$).
The Fourier component is $\bm S_{\bm 0}=\bm0$ in this case, and hence, the electronic state is independent of $J_A^0$ and $J_B^0$.
The $J_A$ and $J_B$ dependence of the charge gap at $n=2$ and $\sigma_{xy}$ is shown in Fig.~\ref{fig:AFM}.
The phase diagram consists of two distinct insulating phases separated by a metallic region; the metallic phase at $|J_A|\gtrsim|J_B|$ is a trivial insulator with the Chern number $C=0$, whereas the insulator phases near the $J_B$ axis are Chern insulators with $C=\pm4$.

\begin{figure}[tbp]
  \includegraphics[width=\linewidth]{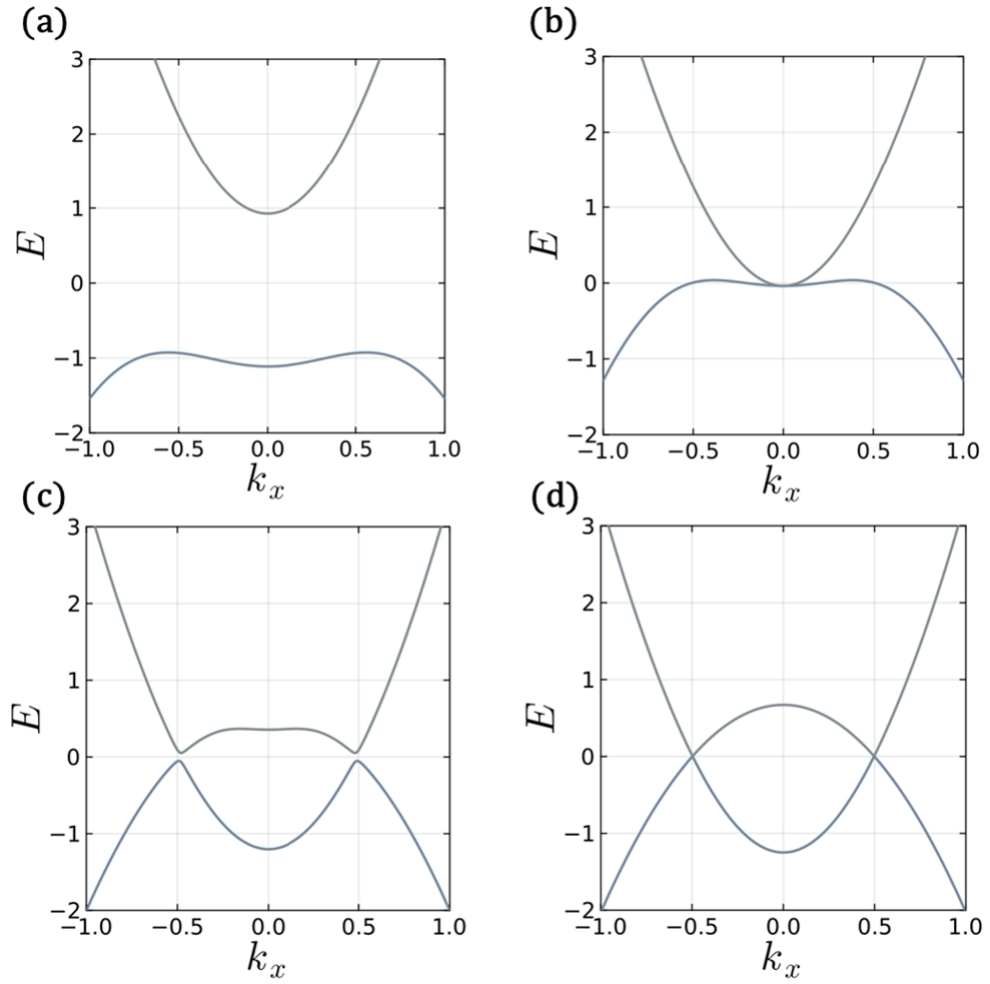}
  \caption{
  Band structure of $H$ for (a) trivial insulator ($J_A=-1.0$, $J_B=-0.5$), (b) double Weyl semimetal ($J_A=-0.48$, $J_B=-0.5$), (c) $C=-4$ Chern insulator ($J_A=-0.1$, $J_B=-0.5$), and (d) the nodal-line-like state ($J_A=0$, $J_B=-0.5$) The results are for $\alpha_0=-2.7$, $\alpha_x= 8.0$, $\alpha_y= 2.7$, $\Delta=0.04$, filling $n=2$, and $J_{A,B}^0=0$. 
  The chemical potential $\mu$ is set so that the Fermi level is $\varepsilon_F=0$.
  }\label{fig:AFM-topo-transition}
\end{figure}

The band structure for the two insulating states is shown in Figs.~\ref{fig:AFM}(c) and \ref{fig:AFM}(d).
The results show a wine-bottle-like structure, which resembles that of the ARPES experiment~\cite{Okuma2024a}.
The evolution of the electronic band from $J_A=0$ is shown in Fig.~\ref{fig:AFM-topo-transition}(a)-\ref{fig:AFM-topo-transition}(c).
The results show that the two insulator phases are separated by a semimetal state.

To study the nature of the phase transition, we next look into the electronic state in detail.
The electronic state at the $\Gamma$ point for $\theta=\arccos(1/3)$ consists of four pairs of doubly-degenerate states, $\varepsilon_{n\tau}$ ($n=1,\cdots,4$),
\begin{align}
    \varepsilon_{1\tau}=&-\sqrt{16 J_A^2 + \Delta^2} - \mu,\\
    \varepsilon_{2\tau}=&-4 J_B + \Delta - \mu,\\
    \varepsilon_{3\tau}=&4 J_B + \Delta - \mu,\\
    \varepsilon_{4\tau}=&\sqrt{16 J_A^2 + \Delta^2} - \mu,
\end{align}
where $\tau=1,2$ is the binary index for the doubly degenerate bands.
The $n=1$ bands are the valence band, and the $n\ge2$ bands are the conduction bands.
Among the conduction bands, the band inversion occurs between the valence band and $n=2$ bands when $J_B>0$, whereas $n=3$ bands move below the conduction band when $J_B<0$.
The phase transition occurs at 
\begin{align}
    |J_B|= \frac\Delta4 + \sqrt{J_A^2 + \left(\frac\Delta4\right)^2},\label{eq:border1}
\end{align}
at which the conduction bands touch the valence bands at $\bm k=\bm0$.

For the $J_A>0$ case, $n=1$ bands touch $n=2$ bands at the phase boundary.
Around the phase boundary, the effective Hamiltonian for the four nearly-degenerate bands consists of two double-Weyl Hamiltonians~\cite{Huang2016a,Zhao2022a},
\begin{align}
    H_\text{DW}=&(\beta_0+\frac\beta2\bm k^2-\mu)\sigma_0+(\gamma_0+\frac{\gamma_1}2\bm k^2)\sigma_3\nonumber\\
    &\qquad+v\frac{k_x^2-k_y^2}2\sigma_1-vk_xk_y\sigma_2,\label{eq:HDW}
\end{align}
where
\begin{align}
    \beta_0=&-2 J_B + \frac\Delta2 - \sqrt{ (2J_A)^2 + \left(\frac\Delta2\right)^2},\\
    \beta=&\frac{4J_A^2 (2 \alpha_0 + 3 (\alpha_x + \alpha_y))}{16 J_A^2 +  \Delta \left(\Delta + \sqrt{16 J_A^2 + \Delta^2}\right)}\nonumber\\
    &+\frac{(2 \alpha_0 + \alpha_x + \alpha_y) \Delta \left(\Delta + \sqrt{16 J_A^2 + \Delta^2}\right)}{32 J_A^2 +  2 \Delta \left(\Delta + \sqrt{16 J_A^2 + \Delta^2}\right)},\\
    v=& \frac{|J_A|(\alpha_x - \alpha_y)}{\sqrt{8J_A^2+\Delta\left(\frac\Delta2+\sqrt{(2J_A)^2+\left(\frac\Delta2\right)^2}\right)}},
\end{align}
\begin{align}
    \gamma_0=&-2 J_B + \frac\Delta2 + \sqrt{ (2J_A)^2 + \left(\frac\Delta2\right)^2},\\
    \gamma_1=&-\frac{(2 \alpha_0 - \alpha_x - \alpha_y) (8 J_A^2 + \Delta\left(\Delta + \sqrt{16 J_A^2 + \Delta^2}\right))}{
 32 J_A^2 + 2 \Delta (\Delta + \sqrt{16 J_A^2 + \Delta^2})}.
\end{align}
Here, $\gamma_0=0$ corresponds to the boundary in Eq.~\eqref{eq:border1}.
Similar to the two-dimensional Dirac electron, by changing $\gamma_0$ from negative to positive, the Chern number changes by $\Delta C=2$.
As there are two degenerate double Dirac electrons, the Chern number changes by $2\Delta C=4$.
This topological phase transition across the two double-Weyl electrons explains the phase transition at the boundary in Fig.~\ref{fig:AFM}(b).

In addition to the double Weyl state at the phase boundary, we find that a nodal-line-like state appears at $J_A=0$ and $J_B^2>\Delta^2/4$.
As shown in Fig.~\ref{fig:AFM-topo-transition}(d), along the $J_B$ axis, the conduction band and valence band touch along a line near the Fermi surface.
This crossing, however, is an accidental one in the absence of $J_A$, distinct from the nodal lines in two dimensions which occur by a crossing of two bands with different mirror index~\cite{Jin2020a,Liu2020b}.
Looking closely into the dispersion, the energy of the band crossing slightly changes depending on the angle $\phi=\arctan(k_y/k_x)$.
However, the change is very small, making the entire nodal-line structure located near the Fermi level.

In the case for $J_A\simeq0$ and $J_B\simeq \Delta/2$ ($\Delta>0$), the double Weyl Hamiltonian in Eq.~\eqref{eq:HDW} gives the effective Hamiltonian for understanding the nodal-line.
For the $J_A=0$ case, the Hamiltonian reads
\begin{align}
    H_\text{DW}=&(\beta_0+\frac\beta2\bm k^2-\mu)\sigma_0+(\gamma_0+\frac{\gamma_1}2\bm k^2)\sigma_3.
\end{align}
Hence, when $\gamma_0/\gamma_1<0$, the conduction and valence bands cross at $k=\sqrt{|2\gamma_0/\gamma_1|}$.
In this limit, all crossing points are at the same energy.
Near the crossing point, the effective Hamiltonian reads
\begin{align}
    H_\text{NL}=&(\beta_0-\frac{\beta\gamma_0}{\gamma_1}+\beta\sqrt{\left|\frac{2\gamma_0}{\gamma_1}\right|}\delta k-\mu)\sigma_0+\sqrt{\left|2\gamma_0\gamma_1\right|}\delta k\sigma_3,
\end{align}
where $\delta k=k-\sqrt{|2\gamma_0/\gamma_1|}$.
The nodal line at $\delta k=0$ is lifted with minimal $J_A$ as it introduces the $v$ term, introducing a gap between the valence and conduction bands, separating the $C=4$ phase into two regions, $J_A>0$ and $J_A<0$.

\subsection{Field-induced insulator-to-metal transition} \label{sec:MItransition}

\begin{figure}[tbp]
  \includegraphics[width=\linewidth]{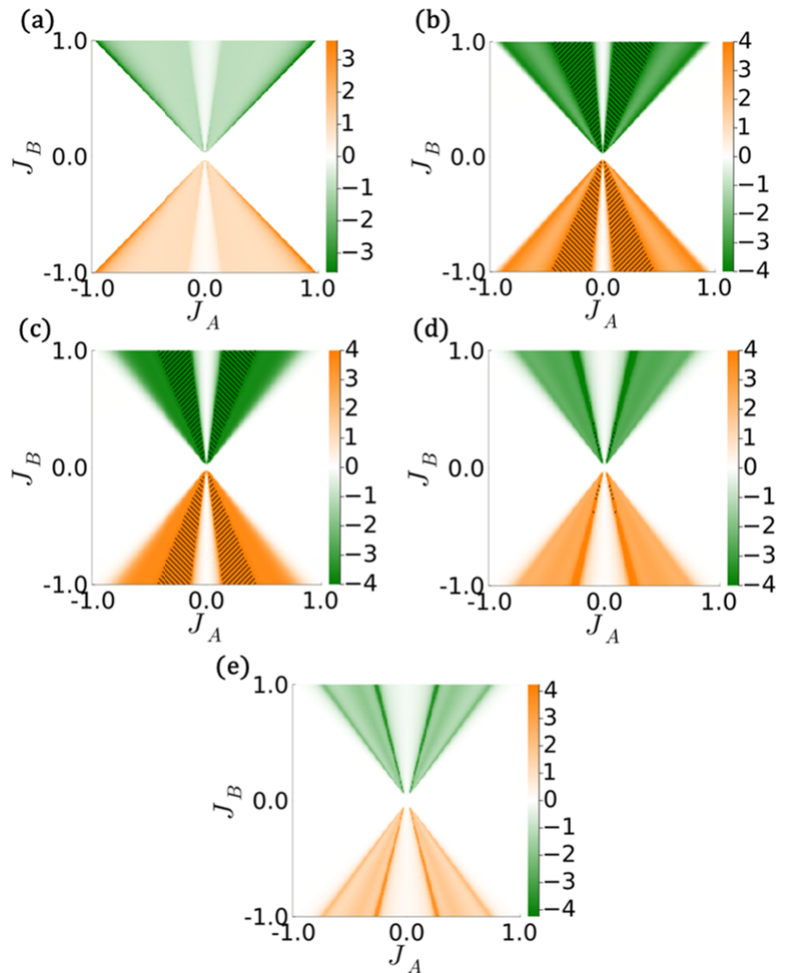}
  \caption{
  Contour plot of $\sigma_{xy}$ for (a) $\pi/6$, (b) $\pi/3$, (c) $\pi/2$, (d) $2\pi/3$, and (e) $5\pi/6$. 
   The hatched regions are Chern insulators.
  The results are for $\alpha_0=-2.7$, $\alpha_x= 8.0$, $\alpha_y= 2.7$, $\Delta=0.04$, filling $n=2$, and $J_{A,B}^0=0$. 
  }\label{fig:theta-dep}
\end{figure}

Next, we look into the $\theta$ dependence of the phase diagram.
Figure~\ref{fig:theta-dep} shows the contour plot of $\sigma_{xy}$ for different canting angles $\theta$.
Here, $\theta=0$ corresponds to the three-up one-down collinear ferrimagnetic state, and $\theta=\pi$ is the ferromagnetic state with all spins pointing down.
The other $\theta$ ($0<\theta<\pi$) are the canted triple-$q$ states.  
The system shows a finite magnetization except for the $\theta=\arccos(1/3)$ discussed in the previous section.
In the experiment, the canting angle is controllable by applying the magnetic field.
Hence, it corresponds to the evolution of electronic states in the magnetic field.

By shifting $\theta$ away from $\arccos(1/3)$, metallic regions appear in between the two insulator phases.
Typically, the metal phases take over the $C=\pm4$ Chern insulator phase near the phase boundary, whereas the trivial insulator region remains robust, with only minor changes.
The $C=\pm4$ insulator phase survives in a range of $\theta$ around $\theta=\arccos(1/3)\simeq0.39\pi$ as shown in the results for $\theta=\pi/3$ and $\pi/2$ in Figs.~\ref{fig:theta-dep}(b) and \ref{fig:theta-dep}(c), respectively.
On the other hand, for larger $\theta$, the Chern insulator is taken over by a metallic phase with a finite topological Hall effect.
At $\theta=0,\pi$, $\sigma_{xy}$ becomes zero for arbitrary $J_{A,B}$ as they are collinear states.
It shows that the $C=\pm4$ state appears over a fairly wide range of canting angles near $\theta=\arccos(1/3)$, provided that $J_B$ is sufficiently large.

The results imply that an insulator-to-metal transition is expected when applying a magnetic field to the Chern insulator phase, in contrast to the trivial insulator, which remains robust under the field.
As stated above, as the canting angle deviates from $\theta=\arccos(1/3)$, the $C=4$ insulator phase is taken over by a metal phase, whereas the trivial insulator phase remains robust for all angles in Fig.~\ref{fig:theta-dep}.
The observation of such an insulator-to-metal transition by the magnetic field provides additional evidence for the topologically nontrivial phase. 

For the $J_{A,B}^0=0$ cases, the $|J_B|\gtrsim |J_A|$ region is completely taken over by the metal phase while the trivial insulator phase remains robust near the colinear ferrimagnetic phase ($\theta=0$).
On the other hand, for the $\theta=\pi$ ferromagnetic state, the entire phase diagram is taken over by the trivial insulator due to $\Delta\;(>0)$; it will be a semimetal if $\Delta<0$.
However, this trend in the collinear state is sensitive to $J_{A,B}^0$ as we will discuss later.

\subsection{$C=2$ Chern insulator}\label{sec:C=2_state}

\begin{figure}[tbp]
  \includegraphics[width=\linewidth]{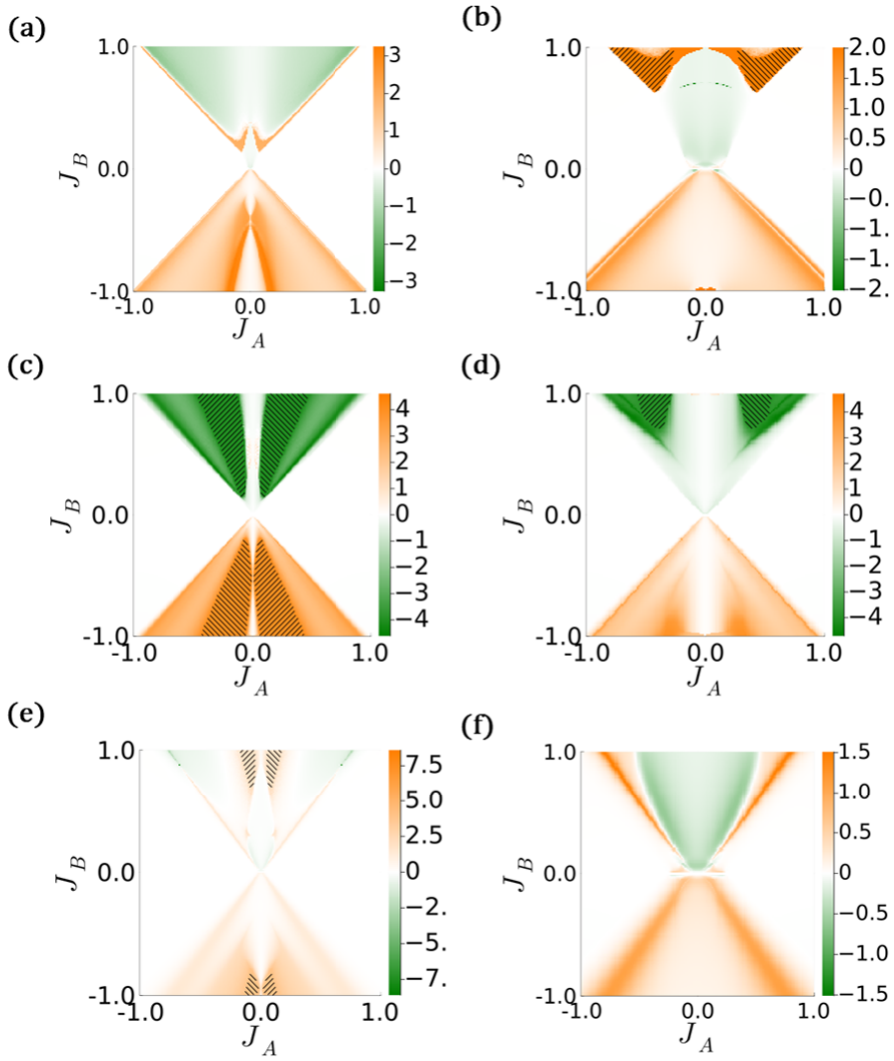}
  \caption{
  Contour plot of $\sigma_{xy}$ for (a) $J_A^0=0.1$ and $\theta=\pi/6$, (b) $J_A^0=0.5$ and $\theta=\pi/6$, (c) $J_A^0=0.1$ and $\theta=\pi/3$, (d) $J_A^0=0.5$ and $\theta=\pi/3$, (e) $J_A^0=0.1$ and $\theta=2\pi/3$, and (f) $J_A^0=0.5$ and $\theta=2\pi/3$.
   The hatched regions are Chern insulators.
  The results are for $\alpha_0=-2.7$, $\alpha_x= 8.0$, $\alpha_y= 2.7$, $\Delta=0.04$, filling $n=2$, and $J_B^0=0$. 
  }\label{fig:JA0}
\end{figure}

When the canting angle slightly deviates from $\theta=\arccos(1/3)$, a small $\bm S_{\bm0}$ term appears; this is likely the case for GdGaI in which a small but finite net magnetization is observed~\cite{Okuma2024a}.
In such cases, $J_A^0$ and $J_B^0$ will also affect the electronic state, in addition to $J_A$ and $J_B$ discussed in Sec.~\ref{sec:JA-and-JB}.
The contour plot of $\sigma_{xy}$ with $J_A^0\ne0$ is shown in Fig.~\ref{fig:JA0}.
Here, we focus on the $J_A^0>0$ region as the contour plot of $J_A<0$ cases is related to the $J_A>0$ cases by time-reversal symmetry; the dispersion is the same, and the Hall conductivity reverses the sign.

A notable feature in the presence of $J_A^0$ is the $C=2$ Chern insulator state that appears at $\theta=\pi/6$ with a sufficiently large $J_B$.
As $J_A^0$ increases, the $C=2$ phase in the $J_B>0$ region enlarges and shares the boundary with the trivial insulator at $J_A^0=0.5$ (Fig.~\ref{fig:JA0}(b)). 
In contrast to the $C=\pm4$ state in the $J_{A,B}^0=0$ case, however, the contour plot is asymmetric on the top and bottom halves.
It is because the Hall conductivity for $(J_A,J_B)$ in Fig.~\ref{fig:AFM} is the opposite of that for $(-J_A,-J_B)$; it is related to Onsager's theorem as reversing the sign of the exchange interactions corresponds to reversing the direction of spins.
For the cases in Fig.~\ref{fig:JA0}, the $\sigma_{xy}$ for $(J_A,J_B,J_A^0)$ is the opposite of $(-J_A,-J_B,-J_A^0)$.
Hence, a $C=-2$ state appears in the $J_B<0$ region at $J_A^0=-0.5$.

For a better understanding of the phase transition between the $C=2$ and trivial insulators, we construct a low-energy theory at the $\Gamma$ point.
With a similar method used in Sec.~\ref{sec:JA-and-JB}, we find that the effective Hamiltonian near the Fermi level is a double Weyl Hamiltonian.
In particular, near $J_A=0.5$, $J_B=0.686$, $J_A^0=0.5$, and $J_B^0=0$, the effective Hamiltonian is $H_{DW}$, with 
\begin{align*}
    v\simeq&\,0.556,\\
    \beta_0\simeq&-1.836 - 0.609\,\delta J^0_A - 0.571 J^0_B\nonumber\\
    &\quad - 1.521\,\delta J_A - 1.138\,\delta J_B,\\
    \beta_1\simeq&\,4.560,\\
    \gamma_0\simeq&\,0.609\,\delta J^0_A - 0.519 J^0_B + 1.521\,\delta J_A - 1.598\,\delta J_B,\\
    \gamma_1\simeq&\,6.140.
\end{align*}
The result shows that the boundary between the trivial and $C=2$ Chern insulators is a double-Weyl semimetal.
A difference from the $C=\pm4$ case is that there is only one double Weyl node at the Fermi level due to the lack of double degeneracy, which results in a $\Delta C=2$ jump at the transition.
Around the boundary, positive $\delta J_A^0$ and $\delta J_A$, and negative $\delta J_B^0$ and $\delta J_B$, make the system a $C=2$ Chern insulator, and the other side is the trivial insulator phase.

With increasing $\theta$, the $C=\pm4$ Chern insulator appears around $\theta=\arccos(1/3)$, as shown in Figs.~\ref{fig:JA0}(c) and \ref{fig:JA0}(d).
With increasing $J_A^0$, the Chern insulator state generally shrinks, but still remains at $J_A^0=0.5$ in the $J_B>0$ region. 
At $\theta=2\pi/3$, another $C=2$ Chern insulator appears at $J_A^0=0.1$, as shown in Fig.~\ref{fig:JA0}(e).
Hence, by increasing $\theta$ from $0$ to $\pi$ in the large $J_B$ region, we find a phase transition from $C=4$ state to $C=2$ state, and then to the metallic phase near the ferromagnetic state ($\theta=\pi$).

\begin{figure}[tbp]
  \includegraphics[width=\linewidth]{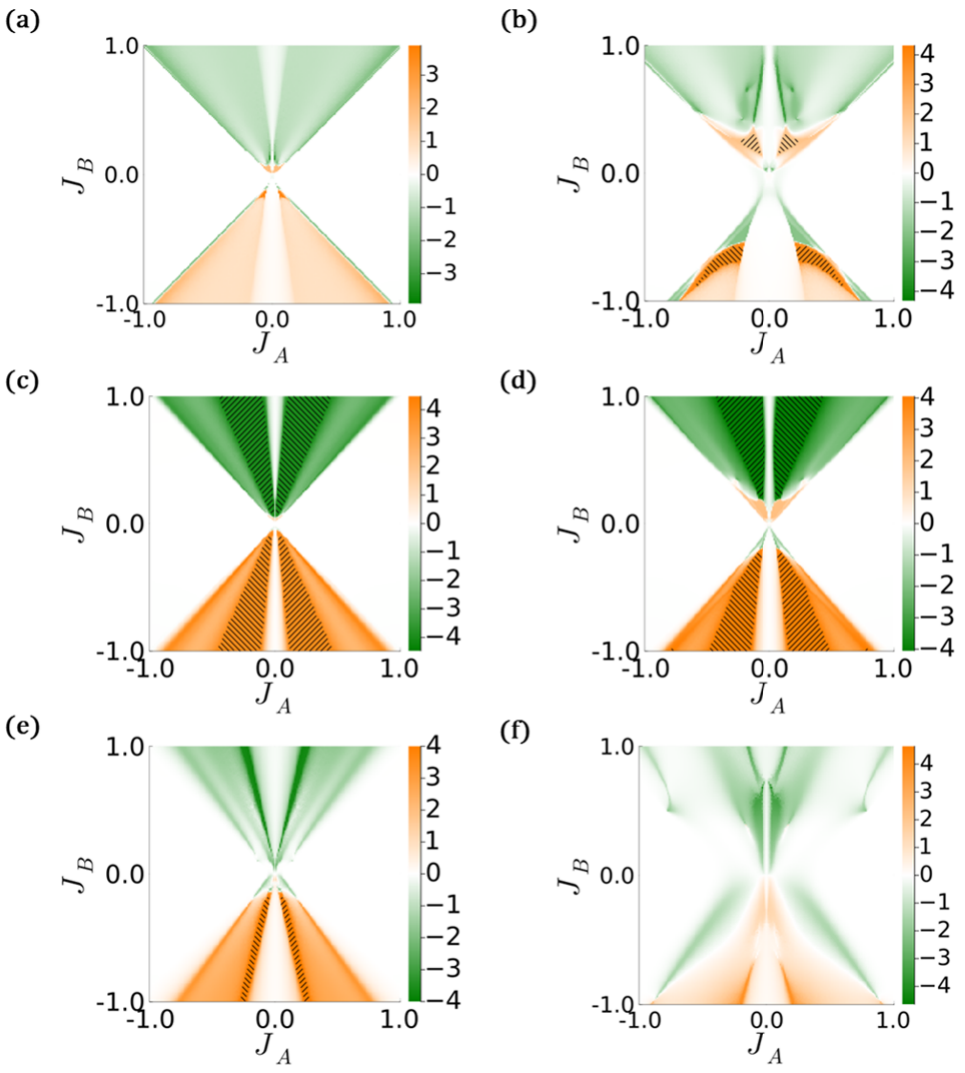}
  \caption{
  Contour plot of $\sigma_{xy}$ for (a) $J_B^0=0.1$ and $\theta=\pi/6$, (b) $J_B^0=0.5$ and $\theta=\pi/6$, (c) $J_B^0=0.1$ and $\theta=\pi/3$, (d) $J_B^0=0.5$ and $\theta=\pi/3$, (e) $J_B^0=0.1$ and $\theta=2\pi/3$, and (f) $J_B^0=0.5$ and $\theta=2\pi/3$.
   The hatched regions are Chern insulators.
  The results are for $\alpha_0=-2.7$, $\alpha_x= 8.0$, $\alpha_y= 2.7$, $\Delta=0.04$, filling $n=2$, and $J_A^0=0$. 
  }\label{fig:JB0}
\end{figure}

A similar trend is found for the $J_B^0\ne0$ case shown in Fig.~\ref{fig:JB0}.
A $C=2$ Chern insulator appears in the $J_B>0$ region at $J_B^0=0.5$ and $\theta=\pi/6$, as in Fig.~\ref{fig:JB0}(b).
In addition, $C=4$ appears in the $J_B<0$ region for a sufficiently large $J_{B}^{0}$ ($J_{B}^{0}=0.5$) (See Figs.~\ref{fig:JB0}(a) and \ref{fig:JB0}(b)).
Near $\theta=\arccos(1/3)$, the $C=\pm4$ Chern insulator survives robustly with $J_B^0$ as demonstrated by the results for $\theta=\pi/3$ in Figs.~\ref{fig:JB0}(c) and \ref{fig:JB0}(d).
For the case of $J_{B}^{0}\neq0$, the $C=4$ state also survives at $\theta=2/3\pi$ for a weak $J_{B}^{0}$($J_{B}^{0}=0.1$) (See Figs.~\ref{fig:JB0} and \ref{fig:JB0}(f)).
The results show that the $C=\pm4$ Chern insulator phase appears in a wide range of $\theta$ in the presence of $J_B^0$, and a $C=2$ Chern insulator appears near the ferrimagnetic state ($\theta=\pi/6$).

Overall, we find that the $C=\pm4$ Chern insulator appears robustly near $\theta=\arccos(1/3)$.
This is natural as, near $\theta=\arccos(1/3)$, the $J_{A,B}^0$ terms are relatively small due to the small spin structure factor at $\bm k=\bm0$, which corresponds to the ferromagnetic moment.
On the other hand, the electronic phase diagram is strongly affected by $J_{A,B}^0$ at the canting angle away from $\arccos(1/3)$, in which a large magnetization exists.
Notably, for a sufficiently large $J_A^0$, a new Chern insulator state with $C=2$ appears at $\theta=\pi/6$.
In contrast, the trivial insulator region is insensitive to $J_{A,B}^0$ as shown in Figs.~\ref{fig:JA0} and \ref{fig:JB0}.
The results in Figs.~\ref{fig:theta-dep}, \ref{fig:JA0}, and \ref{fig:JB0} show that the $C=\pm4$ Chern insulator is generally stable at and around $\theta=\arccos(1/3)$ and is sensitive to the canting angle, showing an insulator-to-metal transition by changing the canting angle, which can be controlled by applying a magnetic field.

\begin{figure*}[tbp]
  \includegraphics[width=\linewidth]{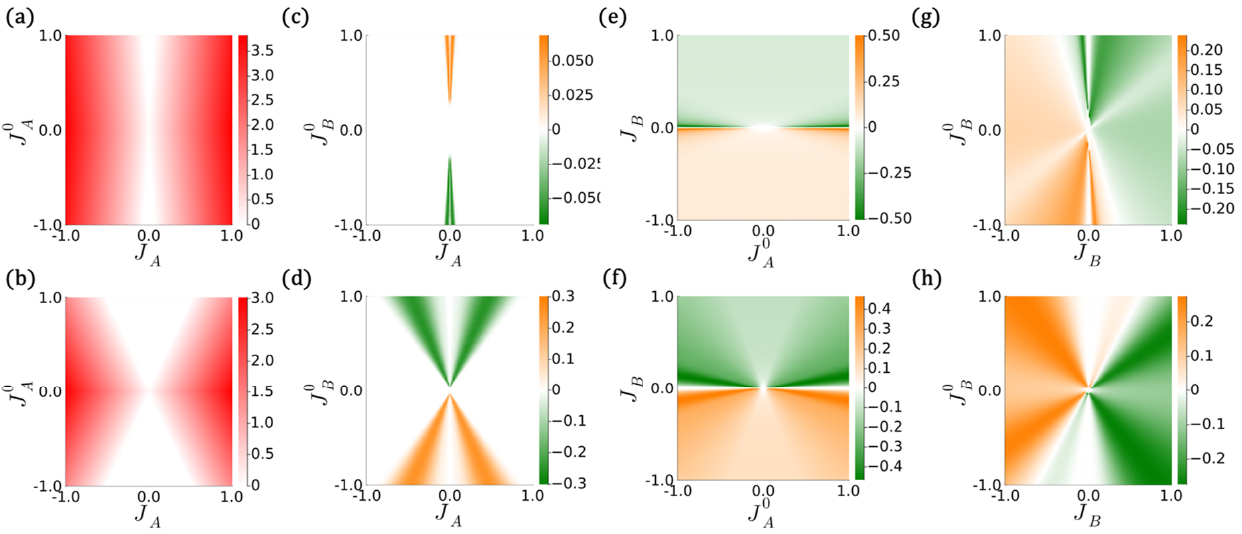}
  \caption{
  Contour plot of band gap for $J_A$ and $J_A^0$ with $J_B=J_B^0=0$ for (a) $\theta=\pi/3$ and (b) $\theta=2\pi/3$, and $\sigma_{xy}$ for $J_A$ and $J_B^0$ with $J_B=J_A^0=0$ for (c) $\theta=\pi/3$ and (d) $\theta=2\pi/3$, $J_B$ and $J_A^0$ with $J_A=J_B^0=0$ for (e) $\theta=\pi/3$ and (f) $\theta=2\pi/3$, and $J_B$ and $J_B^0$ with $J_A=J_A^0=0$ for (g) $\theta=\pi/3$ and (h) $\theta=2\pi/3$.
  The results are for $\alpha_0=-2.7$, $\alpha_x= 8.0$, $\alpha_y= 2.7$, $\Delta=0.04$, and filling $n=2$.
  }\label{fig:2dim-phasediagrams}
\end{figure*}

\subsection{Metal states in weak $J_A$ regions}\label{sec:metal}

In the last, we look into the cases with a large $J_A^0$ and $J_B^0$.
To this end, we investigated the two-dimensional phase diagrams with $J_A^0$ and $J_B^0$. 
Figure~\ref{fig:2dim-phasediagrams} shows the contour plots of $\sigma_{xy}$ for different pairs of exchange interactions.
The trivial insulator state appears in the regions where $J_A$ is the strongest exchange interaction, as shown in Figs.~\ref{fig:2dim-phasediagrams}(a)-\ref{fig:2dim-phasediagrams}(d).
On the other hand, metal phases appear when $J_A=0$ or weak compared to the other exchange interactions, as in Figs.~\ref{fig:2dim-phasediagrams}(e)-\ref{fig:2dim-phasediagrams}(h).

The absence of insulator states in the weak $J_A$ region is related to hybridization between the valence and conduction bands.
The hybridization does not occur in the absence of $J_A$, as implied from the Hamiltonian in Eq.~\eqref{eq:HK}.
As a consequence, when $\Delta$ is small, a magnetic order tends to prefer a metal state by shifting the valence and conduction bands; one of the valence bands is pushed upward due to the Zeeman coupling, while the Zeeman coupling and hybridization between the $d$ orbital pockets push part of the conduction bands lower.
Hence, a metallic state is stabilized provided that $\Delta$ is small compared to the exchange interactions.
In contrast, in the presence of $J_A$, it induces hybridization between the conduction and valence bands, often resulting in an anti-crossing gap.
If the anti-crossing induces a large gap, which is the case for strong $J_A$, the system turns into an insulator.
The results in Fig.~\ref{fig:2dim-phasediagrams} confirm the argument that strong $J_A$ is essential for realizing the insulators, both trivial and topological.

The above argument also implies that a metallic state is expected in the ferromagnetic state $(\theta=\pi)$.
In the ferromagnetic state, $\bm S_{\bm k_{1,2,3}}$ vanishes, and hence, $J_{A,B}$ does not affect the electronic state.
Physically, it is a consequence of the shorter translation periodicity of the magnetic order; Bloch's theorem prohibits hybridization between different bands.
As a consequence, the ferromagnetic state is expected to be metallic in the presence of $J_{A,B}^0$, as confirmed in Secs .~\ref{sec:C=2_state} and \ref{sec:metal}.

\section{Summary and Discussions}\label{sec:summary}

In summary, we have constructed an effective $\bm{k}\cdot\bm{p}$ model for GdGaI with the four-sublattice triple-$q$ noncoplanar order and clarified its electronic and topological properties.
For the antiferromagnetic umbrella state with zero net magnetization ($\theta = \arccos(1/3)$), the phase diagram in the $J_A$-$J_B$ plane hosts a trivial ($C=0$) insulator and $C=\pm 4$ Chern insulators at filling $n=2$, separated by metallic regions. 
Using a low-energy theory at the $\Gamma$ point, we showed that the phase boundary between the trivial and $C=\pm 4$ insulators is described by two degenerate double-Weyl semimetals, naturally explaining the $\Delta C = 4$ jump in the Chern number.
At $J_A=0$ and $|J_B| > \Delta/2$, the valence and conduction bands cross along a line, forming a dispersion similar to a nodal line pinned near the Fermi level, dividing the $C=\pm4$ Chern insulator states into two; this nodal line is gapped out by any finite $J_A$ through the $p$-$d$ hybridization it induces.
We further found that tuning the canting angle $\theta$, controllable by an external magnetic field, drives an insulator-to-metal transition; as $\theta$ deviates from $\arccos(1/3)$, the $C=\pm 4$ Chern insulator phase is gradually taken over by metallic regions, whereas the trivial insulator phase remains robust.
Away from $\theta = \arccos(1/3)$, where the exchange couplings $J_A^0$ and $J_B^0$ associated with the uniform magnetization become non-negligible, a $C=\pm 2$ Chern insulator phase emerges.
Finally, we showed that a sufficiently strong $J_A$ is essential for stabilizing the insulating phases since the $p$-$d$ hybridization it provides is required to gap out the bands inverted by the exchange splitting.

A recent experiment~\cite{Okuma2024a} reports that GdGaI is an insulator with a narrow band gap.
This is roughly consistent with a recent first-principles study, which finds the top of the valence band and the bottom of the conduction band at almost the same energy~\cite{Kaneko2026a}.
While the precise size of $\Delta$, which quantifies the energy difference between the band top and bottom, remains to be confirmed, the existing results and the importance of $J_A$ discussed in Sec.~\ref{sec:metal} imply that $J_A$ is considerably large and plays an important role in stabilizing the insulator state reported in experiment.
In such cases, an insulator-to-metal transition is expected in the magnetic field, as the system approaches the field-forced ferromagnetic phase.

Lastly, we find that a couple of Chern insulator phases appear in the four-sublattice order in large $J_B$ regions; a Chern insulator with $C=\pm4$ is expected to appear in the four-sublattice order, whereas a $C=\pm2$ state is expected near the ferrimagnetic state.
Note that the $C=4$ Chern insulator is also predicted in a recent preprint studying a similar model~\cite{Vylet2026a}.
The $C=\pm4$ Chern insulator state is stable near $\theta=\arccos(1/3)$ as discussed in Secs.~\ref{sec:JA-and-JB}-\ref{sec:C=2_state}.
Hence, possibly appear in GdGaI or in similar compounds if $|J_B|>|J_A|$.
On the other hand, the $C=2$ state may be stabilized by introducing a uniaxial anisotropy, which favors $\theta<\arccos(1/3)$.

\acknowledgements
We are grateful for the fruitful discussions with C. D. Batista, T. Ikenobe, Y. Okada, R. Okuma, and H. Suwa.
This work was supported by JSPS KAKENHI (Grant Numbers JP23K03275, JP25H00841) and JST PRESTO (Grant No. JPMJPR2452).

%
%
%
%
\bibliography{ref} 

\end{document}